\documentclass[journal,twoside,web]{ieeecolor}
\usepackage{generic}
\usepackage{amsmath,amssymb,amsfonts}
\usepackage{algorithmic}
\usepackage{graphicx}
\usepackage{algorithm,algorithmic}
\usepackage{hyperref}
\usepackage{textcomp}
\usepackage[capitalise]{cleveref}
\usepackage{placeins}
\usepackage{subfig}
\usepackage{stfloats}
\usepackage{booktabs}
\usepackage{adjustbox}
\usepackage[backend=biber, style=ieee, url=false, doi=false, maxbibnames=6, minnames=1]{biblatex}
\AtEveryBibitem{\clearfield{issn}}  % Clear ISSN field for every bibliography item

\def\BibTeX{{\rm B\kern-.05em{\sc i\kern-.025em b}\kern-.08em
    T\kern-.1667em\lower.7ex\hbox{E}\kern-.125emX}}
    
\begin{document}
\title{GONet: A Generalizable Deep Learning Model for Glaucoma Detection}

\author{Or Abramovich, Hadas Pizem, Jonathan Fhima, Eran Berkowitz, Ben Gofrit, Meishar Meisel, Meital Baskin, Jan Van Eijgen, Ingeborg Stalmans, Eytan Z. Blumenthal and Joachim A. Behar \IEEEmembership{Senior Member, IEEE}
\thanks{OA, JB, EB and HP acknowledge the support of the Technion EVPR Fund: Irving \& Branna Sisenwein Research Fund. This research was supported by a cloud computing grant from the Israel Council of Higher Education, awarded by the Israel Data Science Initiative. We acknowledge the assistance of ChatGPT, an AI-based language model developed by OpenAI, in editing the manuscript.}
\thanks{Authors’ contribution: JB conceived and designed the research. OA developed the algorithms and performed the analysis under the supervision of JB. JF contributed to the development of the LUNet model for CDR estimation. EZB and HP provided medical guidance throughout the study. JB and OA drafted the first version of the manuscript. JVE and IS contributed and curated the KULRD dataset, refined the inclusion/exclusion criteria for this dataset, and provided medical guidance throughout the study. EB, MM, and MD contributed and curated the HYRD dataset. BG implemented the algorithms on the Lirot.ai iOS platform for public release. All authors discussed the results and edited, revised, and approved the final version of the manuscript}
\thanks{OA, BG and JAB (e-mail: jbehar@technion.ac.il) are affiliated with the Faculty of Biomedical Engineering, Technion, Israel Institute of Technology, Haifa, 3200003, Israel. JF is affiliated with the Department of Applied Mathematics and the Faculty of Biomedical Engineering, Technion, Israel Institute of Technology, Haifa, 3200003, Israel. 
EB, MM, and MB are affiliated with the Hillel Yaffe Medical Center, Hadera, Israel. JVE and IS are affiliated with the Research Group Ophthalmology, Department of Neurosciences, KU Leuven, and with the Department of Ophthalmology, University Hospitals UZ Leuven, Herestraat 49, 3000 Leuven, Belgium. EZB and HP are affiliated with the Rambam Medical Center: Rambam Health Care Campus, Haifa, Israel}
}
\maketitle

%%%%%%%%%%%%%%%%%%%% 
\begin{abstract}
Glaucomatous optic neuropathy (GON) is a prevalent ocular disease that can lead to irreversible vision loss if not detected early and treated. The traditional diagnostic approach for GON involves a set of ophthalmic examinations, which are time-consuming and require a visit to an ophthalmologist. Recent deep learning models for automating GON detection from digital fundus images (DFI) have shown promise but often suffer from limited generalizability across different ethnicities, disease groups and examination settings. To address these limitations, we introduce GONet, a robust deep learning model developed using seven independent datasets, including over 119,000 DFIs with gold-standard annotations and from patients of diverse geographic backgrounds. GONet consists of a DINOv2 pre-trained self-supervised vision transformers fine-tuned using a multisource domain strategy. GONet demonstrated high out-of-distribution generalizability, with an AUC of 0.85-0.99 in target domains. GONet performance was similar or superior to state-of-the-art works and was significantly superior to the cup-to-disc ratio, by up to 21.6\%. GONet is available at [URL provided on publication]. We also contribute a new dataset consisting of 768 DFI with GON labels as open access.

\end{abstract}


%%%%%%%%%%%%%%%%%%%%
\section{Introduction}
\label{sec:introduction}

\IEEEPARstart{G}{laucomatous} optic neuropathy (GON) is a leading cause of irreversible blindness worldwide \cite{Gupta2016PrevalenceSurvey}. It is characterized by damage to the retinal ganglion cells, the retinal nerve fiber layer and the optic nerve, leading to permanent vision loss and eventually to blindness \cite{Gupta2016PrevalenceSurvey}. GON is incurable, but early detection and treatment can stop or at least slow the progression of the disease and reduce the risk of severe vision loss. In 2013, 64.3 million people worldwide between the ages of 40 and 80 years had GON, with the number of affected individuals expected to reach 111.8 million by 2040 \cite{Gupta2016PrevalenceSurvey} \cite{Tham2014GlobalMeta-analysis}. Approximately 50\% of all cases of GON are undiagnosed, mainly because symptoms, such as vision loss, are first noticed when the disease is already at an advanced stage \cite{Stevens2013Global1990-2010}. 
%In 2021, the overall economic burden of visual impairment and blindness worldwide, including direct medical costs and indirect costs, such as productivity losses and long-term disability, was estimated to be \$322.1-518.7 billion \cite{Marques2021GlobalBlindness}. 

GON is diagnosed through a comprehensive ophthalmic examination that includes inspection of the optic disc (OD), imaging of the optic nerve head, and visual field assessment \cite{Gupta2016PrevalenceSurvey, Spaeth2021EuropeanGlaucoma}. While these examinations effectively detect GON, they require the expertise of an ophthalmologist and access to specialized, often costly, equipment which can be a limiting factor. Alternatively, computer-aided analysis of digital fundus images (DFI) can be used to identify GON. DFIs are captured using a fundus camera, which photographs the posterior segment of the eye and provides a clear view of the OD \cite{Abramovich2023FundusQ-Net:Grading}. The OD features a central, depressed area known as the cup, surrounded by the neuroretinal rim, composed of nerve fibers that converge to form the optic nerve. In GON, loss of these fibers causes the rim to shrink and the cup to enlarge \cite{Schuman2020ReviewLecture}, while the overall size of the disc remains unchanged. Thus, the cup-to-disc ratio (CDR), defined as the vertical ratio of the diameter of the cup versus the diameter of the disc, serves as a critical indicator of the presence and severity of GON \cite{Gupta2016PrevalenceSurvey}. However, CDR can be affected by the natural variations in OD size between individuals \cite{Morgan2005DigitalAnalysis}. Furthermore, the CDR by itself may not capture all the structural changes associated with GON \cite{Fhima2024ComputerizedGlaucoma, Hemelings2021DeepDisc}. Recent works have presented alternative features that may better capture signs of GON, such as the rim-to-disc ratio (RDR) \cite{Kumar2019Rim-to-DiscPrescreening}. Finally, clinical CDR estimation varies between experts and between observations \cite{Spaeth2021EuropeanGlaucoma, Morgan2005DigitalAnalysis}. Taken together, usage of CDR for GON identification is prone to misclassification.

%%%%%
\begin{figure*}[t]
    \centering
    \includegraphics[width=1\textwidth]{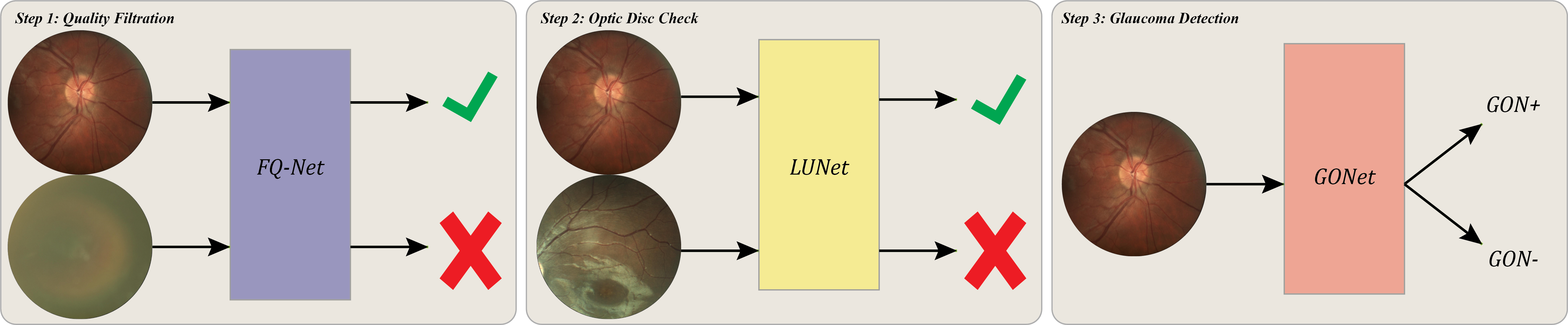}
    \caption{Summary of the proposed research. We propose an end-to-end pipeline for the identification of GON from digital fundus images (DFI), based on a new deep learning model denoted GONet. Panel A: Low quality DFI filtration via FundusQ-Net \cite{Abramovich2023FundusQ-Net:Grading}; Panel B: Missing OD filtration via LUNet \cite{Fhima2024LUNet:Images}; Panel C: GON identification using GONet.}
    \label{fig:pipeline}
\end{figure*}

%%%%%%%%%%%%
A significant body of literature has been published on the diagnosis of GON using DFIs \cite{Thompson2020AProgression}, with recent studies increasingly using deep learning (DL) for GON detection \cite{Zedan2023AutomatedReview, Bali2024AnalysisReview}. However, the published studies have notable limitations. Often the GON reference labels were derived solely from DFI evaluations rather than from comprehensive ophthalmic examinations \cite{Christopher2018PerformancePhotographs, Fu2018Disc-AwareImage}, \cite{ Li2018EfficacyPhotographs, Liu2019DevelopmentPhotographs}. This approach intrinsically reduces the GON detection task to a subjective evaluation of the OD, which has inherent limitations in identifying GON. Hence, reliance of DL models solely on DFIs for GON identification can inadvertently lead to biases in the model training. Indeed, such models can inherit a skewed representation of GON, shaped by subjective interpretations and inherent assumptions of DFI annotators, thereby potentially diverging from the accurate clinical manifestation of the condition. In this study, these reference labels are referred to as ``silver-standard'', while gold-standard annotations are those provided by an ophthalmologist, based on a comprehensive examination \cite{Camara2022RetinalMissing}. In addition, a large proportion of GON diagnostic research with DL models failed to evaluate the model performance on external datasets \cite{Christopher2018PerformancePhotographs,  Chakravarty2018AImages, Li2018EfficacyPhotographs}. Few studies evaluated model generalizability with external datasets \cite{Liu2019DevelopmentPhotographs, Fu2018Disc-AwareImage}, and those that did, reported on a significant drop in performance on external datasets, underscoring limitations in out-of-distribution generalization. 
The works of Hemelings et al. \cite{Hemelings2021DeepDisc, Hemelings2023AImages} focused on generalization, introducing a ResNet-based model, G-RISK, trained on 16,799 DFIs from the KU Leuven Hospital in Belgium. The generalization performance of G-RISK was evaluated on 13 external datasets, 7 of which had gold standard annotations. When using a constant probability decision threshold, out-of-distribution (OOD) generalization performance ranged 0.70-0.98 sensitivity and 0.74-0.94 specificity for the datasets with gold standard annotations and 0.74-0.96 sensitivity and 0.68-0.94 specificity for the datasets with silver standard annotations.

%However, when tested across 13 external datasets, G-RISK performance varied, and required dataset-specific decision thresholds ranging from 0.58 to 0.82 to achieve optimal results. When using a uniform threshold of 0.7, 5 of the 13 datasets reported sensitivity or specificity below 0.75, and as low as 0.68 thus highlighting substantial domain shifts.

Taken together, there is an urgent need for a universal GON diagnostic model that both excels on a local test set and remains effective in diverse populations and examination settings. To address these gaps, this research introduces a novel approach that combines self-supervised learning (SSL) and multi-source-domain (MSD) fine-tuning to develop GONet, a highly generalizable DL model for GON identification. The experiments were performed using seven independent DFI datasets with gold-standard annotations. 

The manuscript starts with a detailed overview of the datasets employed in the experiments. A comparative analysis of various state-of-the-art vision transformer architectures pre-trained using SSL or supervised learning is then conducted to select the most suitable one. Subsequently, the value of MSD training on improving OOD generalization performance is evaluated. The performance of the final model, denoted GONet, is compared against CDR and the RDR \cite{Kumar2019Rim-to-DiscPrescreening}. This research work provides the following three main scientific contributions:
\begin{itemize}
    \item Benchmark of state-of-the-art vision transformer architectures pre-trained using SSL or supervised learning for the task of GON classification.
    \item GONet, a DL model with high OOD generalization for GON identification from a single DFI.
    \item A new open-access dataset, denoted HYRD (288 patients and 768 DFIs), of DFIs with gold-standard GON labels.
    %\item Thorough error analysis investigating the main source of errors of the model.
\end{itemize}

% The manuscript concludes with an in-depth error analysis to elucidate the reasons behind misclassifications.

\begin{figure*}[]
\centering
\noindent\textbf{Table I: Datasets used for the experiments}
\vspace{0.5em}
% Table

\resizebox{\textwidth}{!}{%
    \begin{tabular}{@{}ccclcccccccc@{}}
    \toprule
\textbf{Dataset} & \textbf{\#Patients} & \textbf{\begin{tabular}[c]{@{}c@{}}\#DFIs\\ Total\end{tabular}} & \multicolumn{1}{c}{\textbf{}} & \textbf{\begin{tabular}[c]{@{}c@{}}\#DFIs\\ Selected\end{tabular}} & \textbf{GON+ (\%)} & \textbf{Geography} & \textbf{Male (\%)} & \textbf{Age (Years)} & \textbf{Camera} & \textbf{FOV (°)} & \textbf{Resolution} \\ \midrule
KULRD & 12,071 & 115,668 & $\Rightarrow$ & 61,213 & 70 & Belgian & 48 & 2-99 & Visucam 500 (Zeiss) & 30 & 1444x1444 \\
HYRD & 288 & 768 & $\Rightarrow$ & 647 & 74 & Israeli & 50 & 36-95 & DRI OCT Triton (Topcon) & 45 & \begin{tabular}[c]{@{}c@{}}2576x1934 \\ 1960x1934\end{tabular} \\
PAPILA \cite{Kovalyk2022PAPILA:Assessment} & 244 & 488 & $\Rightarrow$ & 393 & 18 & Spanish & 62 & 15-90 & TRC-NW400 (Topcon) & 30 & 2576x1934 \\
DRISHTI-GS \cite{Sivaswamy2014Drishti-GS:Segmentation} & - & 101 & $\Rightarrow$ & 89 & 69 & Indian & 50 & 40-80 & - & 30 & 2047x1760 \\
REFUGE \cite{Orlando2020REFUGEPhotographs} & - & 1,200 & $\Rightarrow$ & 1,199 & 10 & Chinese & 47 & - & \begin{tabular}[c]{@{}c@{}}Visucam 500 (Zeiss) \\ CR-2 (Canon)\end{tabular} & - & \begin{tabular}[c]{@{}c@{}}2124x2056\\  1634x1634\end{tabular} \\
REFUGE2 \cite{Fang2022REFUGE2Screening} & - & 800 & $\Rightarrow$ & 796 & 20 & Chinese & - & - & \begin{tabular}[c]{@{}c@{}}TRC-NW400  (Topcon)\\ KOWA\end{tabular} & \begin{tabular}[c]{@{}c@{}}30\\ 45\end{tabular} & \begin{tabular}[c]{@{}c@{}}1848x1848\\ 1940x1940\end{tabular} \\
GAMMA \cite{Wu2023GAMMAImAges} & 276 & 300 & $\Rightarrow$ & 299 & 50 & Chinese & 58 & 19-77 & \begin{tabular}[c]{@{}c@{}}TRC-NW400  (Topcon)\\ KOWA\end{tabular} & \begin{tabular}[c]{@{}c@{}}30\\ 45\end{tabular} & \begin{tabular}[c]{@{}c@{}}1934x1956 \\ 2000x2992\end{tabular} \\ \bottomrule
\end{tabular}%
}\captionsetup{labelformat=empty}
\captionof{table}{GON+ (\%): the prevalence of GON+ DFI in the original dataset (i.e., before applying exclusion criteria). \#DFI Selected: the data subset used after applying exclusion criteria. FOV: the field of view of the fundus camera. Sex prevalence, age range are provided for the original dataset.}
\label{tab:external-datasets}

%\vspace{1em}

% Figure
%\includegraphics[width=\textwidth]{new_figures/DFI Figure Clean.png}
%\captionsetup{labelformat=default}
%\captionof{figure}{Examples from DFI datasets used in this research. Each row represents a dataset, according to the order of Table \ref{tab:external-datasets}. DFIs from different datasets have various fields of view, resolution, quality and positioning of the optic disk.}
%\label{fig:dfis}

\end{figure*}
%%%%%%%%%%%
\section{Materials and Methods}
\label{sec:materials}
\subsection{Datasets}
Five open DFI datasets were included in the experiment. They were selected according to the following criteria: datasets that had gold-standard annotation; ``uncropped'' DFIs, which refers to DFIs that were not cropped around the OD region;  with at least 100 DFIs and at least 30 GON+ DFIs. In addition, two private datasets were used: a dataset from KU Leuven, denoted KU-Leuven retinal dataset (KULRD, Helsinki approval number S60649), and a dataset from the Hillel Yaffe Hospital, denoted the Hillel Yaffe retinal dataset (HYRD, Helsinki approval number 0029-24-HYMC), which is made open-access via Physionet. For all datasets, the following exclusion criteria were used: children ($<$18 years old), GON suspects, DFIs lacking complete OD (LUNet \cite{Fhima2024LUNet:Images}) and low-quality DFIs (FundusQ-Net$<$5 \cite{Abramovich2023FundusQ-Net:Grading}). However, to maintain a fair comparison with results reported in other studies on open datasets, a separate comparison is performed on those datasets without applying any exclusion, i.e. performance is reported for all DFIs included in the given dataset. Table \ref{tab:external-datasets} summarizes the datasets included in this research
%, and Figure \ref{fig:dfis} presents examples of DFIs from these datasets. 

%%%
\subsubsection{KULRD}
KULRD was established by the KU-Leuven glaucoma clinic, and includes data collected between 2010 and 2019 from 13,249 patients, in the framework of the study  ``Automatic glaucoma detection, a retrospective database analysis” (study number S60649). The dataset contains 115,668 DFIs acquired over 31,429 clinic visits. In 92\% (n=28,916) of the visits, the images were stereoscopic images of both eyes, resulting in a total of four images per visit (Zeiss Visucam 500). The dataset contains 59,997 diagnoses for 12,071 patients, i.e., not every DFI has a paired diagnosis. The diagnoses belong to 1,196 categories, which encompass various diseases, treatments and surgeries. Diagnoses can be for both eyes or for one eye only. All diagnostic codes were reviewed and categorized as GON+ for a positive GON diagnosis/surgery, GON- for a diagnostic code unrelated to GON, GON suspect/ocular hypertension for diagnosis relating to GON suspect/ocular hypertension diagnosis or Unknown when it was impossible to know if the category is related to GON.

Then, patients were labeled according to their diagnoses. Since GON is incurable, once an eye is diagnosed as GON+, every subsequent DFI was also labeled as GON+. While patients with unilateral GON have a greater chance of developing bilateral GON \cite{Niziol2018AssociationParticipants}, it is not mandatory, and, as such, it was decided to keep the label based on the eye and not per patient. Due to the chronic nature of GON, the disease can exist for years before any damage manifests \cite{Schuman2020ReviewLecture}. Therefore, it would be incorrect to label DFIs as GON- if they were taken before a first documented GON-positive diagnosis. Consequently, it was decided to exclude DFIs that preceded a documented GON diagnosis. 
Additionally, ocular hypertension, characterized by a normal optic disc appearance coupled with elevated IOP, was also excluded. This is because ocular hypertension is considered an early sign of GON, with 10\% of people developing GON within 5 years of diagnosis \cite{Gordon2002OHTS}.
Finally, due to the unspecific nature of GON suspect, these DFIs were excluded from the study (Figure \ref{fig:eligibility-flowchart}). A total of 8,203 patients and 61,213 DFIs remained in KULRD after applying the above exclusion criteria. They were divided into 85:5:10 train-validation-test sets, while stratifying by age, sex and GON label.

%%%
\subsubsection{REFUGE} 
The REFUGE dataset consists of 1,200 DFIs collected from 600 Chinese subjects \cite{Orlando2020REFUGEPhotographs}. The DFIs were acquired at the Zhongshan Ophthalmic Center of Sun Yat-Sen University, China, using a Zeiss Visucam 500 (n=400) or the Canon CR-2 (n=800). DFIs classified as GON+ comprise 10\% of the dataset. Additionally, OC and OD segmentations were annotated by seven independent GON specialists, each with over five years of experience \cite{Orlando2020REFUGEPhotographs}. These OC/OD segmentations were used for quality control of the CDR estimation algorithm.

%%%
\subsubsection{REFUGE2}
The REFUGE2 dataset consists of 800 DFIs. Like the original REFUGE dataset, it was acquired by the Zhongshan Ophthalmic Center \cite{Fang2022REFUGE2Screening}. Half were acquired using a KOWA device (n=400), and half were acquired using a TOPCON TRC-NW400 non-mydriatic retinal camera (n=400). GON+ DFIs comprise 20\% of the dataset.

%%%
\subsubsection{PAPILA}
The PAPILA dataset is a comprehensive collection of records from 244 patients, each providing structured information that includes clinical data, DFIs, and optic disc and cup segmentations for both eyes of the same patient \cite{Kovalyk2022PAPILA:Assessment}. Diagnostic labels based on clinical data are also provided, classifying patients into three categories: glaucomatous, nonglaucomatous, or suspect. The DFIs were captured by ophthalmologists or technicians at HGURS (Murcia, Spain), using a Topcon TRC-NW400 non-mydriatic retinal camera. GON suspect DFIs were excluded from this study.

%%%
\subsubsection{DRISHTI-GS}
The DRISHTI-GS dataset comprises a total of 101 DFIs, collected at the Aravind Eye Hospital in Madurai, India. The GON+ patients were selected by clinical investigators during their examinations, while healthy subjects were selected from individuals undergoing routine refraction tests. The subjects included in the dataset spanned an age range of 40-80 years, with an approximately equal distribution of male and female subjects \cite{Sivaswamy2014Drishti-GS:Segmentation}.

%%%
\subsubsection{GAMMA}
The GAMMA dataset includes 300 DFIs from 276 Chinese patients, provided by the Sun Yat-Sen Ophthalmic Center, Sun Yat-Sen University, China. The DFIs were acquired with a KOWA or Topcon TRCNW400 camera. DFIs were manually quality checked. The patients were aged 19 to 77 years, and 42\% were female. 
%The DFIs are labeled as not GON, early GON or intermediate/advanced GON. GON grading was based on the mean deviation values from visual field tests \cite{Wu2023GAMMAImAges}. For the current analysis, only the ``not GON'' label was considered GON-.

%%%
\subsubsection{HYRD}
The HYRD dataset comprises 786 DFIs collected from 288 patients following a comprehensive ophthalmic examination and was specially developed for this study. The data were collected by the Hillel Yaffe Ophthalmology Department Glaucoma Unit, Hadera, Israel. The DFIs were taken using a TOPCON DRI OCT Triton retinal camera. The subjects were aged 36 to 95 years, and 74\% of the DFIs were GON+. %HYRD was accessed under a limited-use agreement that permits the use of the data solely for evaluation purpose.

\begin{figure}[]
    \includegraphics[width=1\columnwidth]{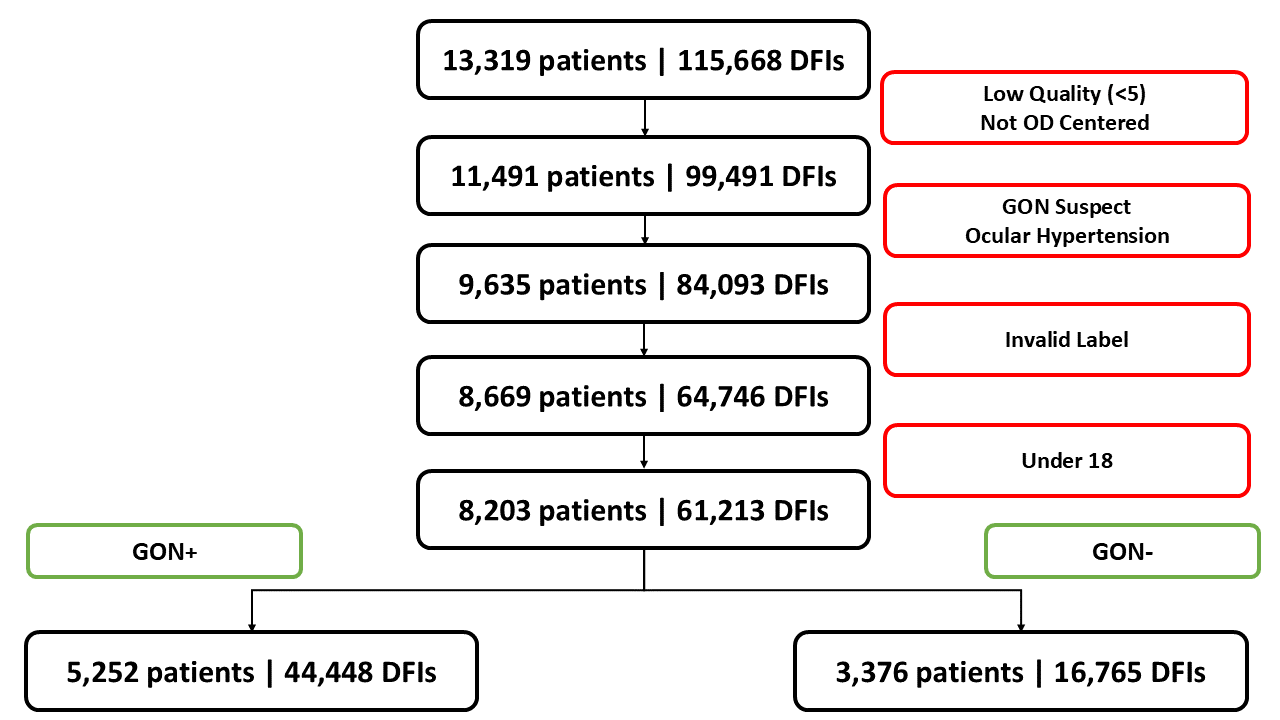}
    \caption{Selection and labeling of the KULRD data. The flowchart summarizes the results of the data selection process.}
    \label{fig:eligibility-flowchart}
\end{figure}

%%%
\subsection{FundusQ-Net for quality estimation}
Real-world DFIs often suffer from poor quality due to various factors, including small pupils, improper flash and gamma adjustments, blinking, as well as media opacity and variations in technician expertise \cite{Abramovich2023FundusQ-Net:Grading}. FundusQ-Net \cite{Abramovich2023FundusQ-Net:Grading} was used to evaluate and filter out low-quality DFIs. FundusQ-Net uses a 1-10 quality grading scale with increments of 0.5, which facilitates quality filtration by setting a defined quality threshold. In this research, following a consultation with two ophthalmologists (EZB and HP), a threshold of 5 was chosen.

%%%
\subsection{LUNet for CDR estimation}
\label{subsec:lunet}
Since capturing the OD is crucial for identifying GON, we excluded DFIs that did not contain the OD. We retrained LUNet \cite{Fhima2024LUNet:Images}, a DFI segmentation model initially developed for retinal vasculature segmentation, for the task of OD segmentation. A publicly available dataset comprising 1,440 DFIs with manual OD/OC segmentations from the G1020 and ORIGA datasets \cite{Bajwa2020G1020:Detection, Zhang2010ORIGA-light:Research} was used for this purpose. The CDR was calculated using LUNet's segmentations of the OC and OD. We validated its performance in estimating CDR on the REFUGE dataset by measuring the mean absolute error (MAE) between the LUNet-predicted and reference CDR values. The performance of LUNet was compared to that reported in the recent work of Gao et al. \cite{Gao2024AutomatedDetection} on REFUGE \cite{Orlando2020REFUGEPhotographs}.
% The resulting model is available as open source within the PVBM toolbox \cite{Fhima2022PVBM:Segmentation} for reproducibility. 

%%%
\subsection{Machine learning}

\subsubsection{Preprocessing}
%DFIs sometimes contain non-fundus black regions in the edges of the DFI \cite{Sivaswamy2014Drishti-GS:Segmentation}. 
To create 1:1 aspect ratio, black padding was added to the DFIs. Next, the DFIs were downsampled to a resolution of 392x392 pixels. Finally, the DFIs were normalized using the mean and standard deviation of the ImageNet dataset \cite{Russakovsky2015ImageNetChallenge}.

\subsubsection{Vision transformers}
Vision transformers have become the popular choice for medical image classification tasks \cite{AzadAdvancesReview}. In this study, we evaluated two specific Vision transformer architectures: ViT-B \cite{Dosovitskiy2020AnScale} and SwinV2 \cite{Liu2022SwinResolution}, with SwinV2 undergoing pre-training through supervised learning. Four state-of-the-art (SOTA)
SSL Vision transformer techniques were benchmarks: DINOv2 \cite{Oquab2024DINOv2:Supervision}, Mugs \cite{Zhou2022Mugs:Framework}, MoCoV3 \cite{Chen2021AnTransformers}, and RETFound \cite{Zhou2023AImages}. All SSL techniques utilized ViT-B as their backbone, maintaining consistent input sizes and training hyperparameters. The models underwent fine-tuning and were subsequently assessed on the KULRD dataset. Notably, all models, except for RETFound which was pre-trained on retinal images, were initially pre-trained on natural images. RETFound stands out as a foundational model pre-trained on 900K DFIs\cite{Zhou2023AImages} using Masked Autoencoder (MAE) \cite{He2021MaskedLearners} as the SSL technique. The most effective architecture and pre-training combination, as determined by performance on the KULRD-Test set, was selected for further experimentation.

% Previous works have reported that using SSL pretraining on DFI for related DFI downstream tasks yields better results \cite{Abramovich2023FundusQ-Net:Grading, Zhou2023AImages}. 
%%%%%%%%%%%%
\begin{figure*}[h]
    \centering
    \includegraphics[width=1\textwidth]{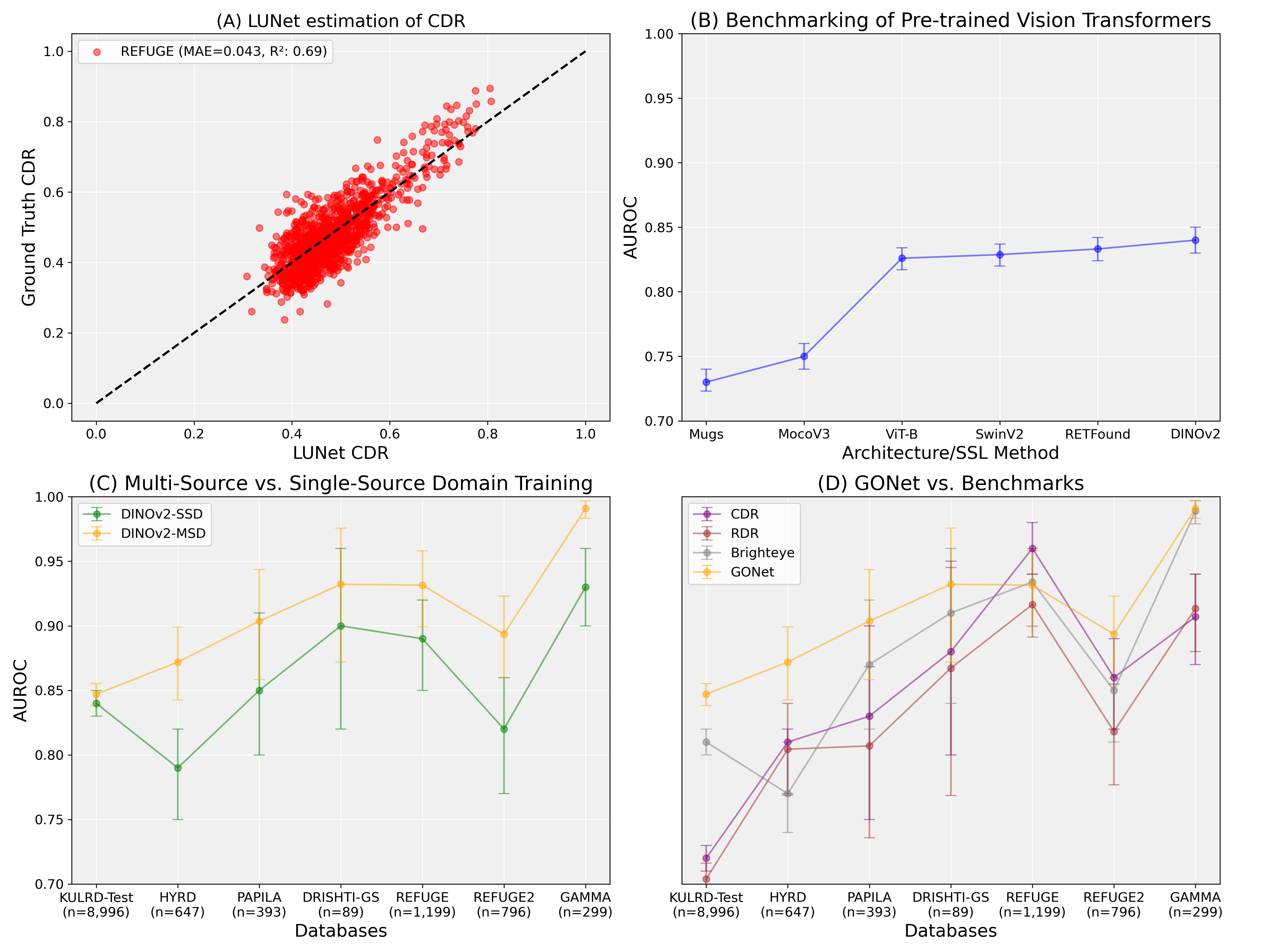}
    \caption{Result figures. Panel A: LUNet estimation of the vertical cup-to-disc ratio (CDR). Results are reported for the REFUGE dataset. Panel B: Performance comparison of alternative pre-trained vision transformers. Fine-tuning to the downstream task was performed on KULRD-Train and performance is reported for KULRD-Test. Panel C: Single-source domain training (SSD) versus multi-source domain training (MSD) strategies for a DINOv2 backbone. Panel D: Performance of GONet (DINOv2 backbone and MSD training) for GON identification versus baselines using the CDR or rim-to-disc ratio (RDR) and a benchmark open-source model called Brighteye \cite{Lin2024Brighteye:Transformer}. For panels B-D, CI was calculated as detailed in section \ref{subsec:performance_measures}.}
    \label{fig:main_results}
    \vspace{2em}
\end{figure*}

\begin{figure*}[h]
\centering
\noindent\textbf{Table II: Results}
\vspace{0.5em}

\resizebox{2\columnwidth}{!}{%
\begin{tabular}{@{}lccccccc@{}}
\toprule
\multicolumn{1}{c}{\textbf{Model}} &
\textbf{\begin{tabular}[c]{@{}c@{}}KULRD-Test (n=8,996)\\ AUC (95\% CI)\end{tabular}} &
\textbf{\begin{tabular}[c]{@{}c@{}}HYRD (n=647)\\ AUC (95\% CI)\end{tabular}} &
\textbf{\begin{tabular}[c]{@{}c@{}}PAPILA (n=393)\\ AUC (95\% CI)\end{tabular}} &
\textbf{\begin{tabular}[c]{@{}c@{}}DRISHTI-GS (n=89)\\ AUC (95\% CI)\end{tabular}} &
\textbf{\begin{tabular}[c]{@{}c@{}}REFUGE (n=1,199)\\ AUC (95\% CI)\end{tabular}} &
\textbf{\begin{tabular}[c]{@{}c@{}}REFUGE2 (n=796)\\ AUC (95\% CI)\end{tabular}} &
\textbf{\begin{tabular}[c]{@{}c@{}}GAMMA (n=299)\\ AUC (95\% CI)\end{tabular}} \\ \midrule
ViT-B (No SSL) & 0.83 (0.82-0.83) & 0.77 (0.74-0.81) & 0.83 (0.78-0.89) & 0.85 (0.77-0.92) & 0.84 (0.80-0.89) & 0.81 (0.77-0.85) & 0.85 (0.84-0.91) \\
Mugs  \cite{Zhou2022Mugs:Framework} & 0.73 (0.72-0.74) & 0.52 (0.47-0.57) & 0.66 (0.59-0.72) & 0.58 (0.44-0.70) & 0.64 (0.58-0.69) & 0.63 (0.58-0.68) & 0.73 (0.67-0.79) \\
Moco-V3 \cite{Chen2021AnTransformers} & 0.75 (0.74-0.76) & 0.36 (0.31-0.41) & 0.73 (0.67-0.78) & 0.58 (0.45-0.70) & 0.65 (0.60-0.71) & 0.64 (0.59-0.69) & 0.70 (0.64-0.76) \\
SwinV2 \cite{Liu2022SwinResolution} & 0.83 (0.82-0.84) & 0.79 (0.75-0.83) & 0.87 (0.82-0.92) & 0.91 (0.85-0.96) & 0.88 (0.85-0.91) & 0.85 (0.81-0.88) & 0.93 (0.91-0.96) \\
RETFound \cite{Zhou2023AImages} & 0.83 (0.82-0.84) & 0.85 (0.82-0.88) & 0.88 (0.84-0.91) & 0.85 (0.77-0.92) & 0.90 (0.86-0.94) & 0.86 (0.82-0.90) & 0.90 (0.87-0.94) \\
DINOv2 \cite{Oquab2024DINOv2:Supervision} & 0.84 (0.83-0.85) & 0.79 (0.75-0.82) & 0.85 (0.80-0.91) & 0.90 (0.82-0.96) & 0.89 (0.85-0.92) & 0.82 (0.77-0.86) & 0.93 (0.90-0.96) \\
RDR (Baseline) \cite{Kumar2019Rim-to-DiscPrescreening} & 0.70 (0.69-0.71) & 0.80 (0.77-0.84) & 0.81 (0.74-0.87) & 0.87 (0.77-0.94) & 0.92 (0.89-0.94) & 0.82 (0.78-0.85) & 0.91 (0.88-0.94) \\
CDR (Baseline) & 0.72 (0.71-0.73) & 0.81 (0.77-0.82) & 0.83 (0.75-0.90) & 0.88 (0.80-0.95) & \textbf{0.96 (0.94-0.98)} & 0.86 (0.82-0.89) & 0.91 (0.87-0.94) \\
Brighteye (Benchmark) \cite{Lin2024Brighteye:Transformer} & 0.81 (0.80-0.82) & 0.77 (0.74-0.81) & 0.87 (0.82-0.92) & 0.91 (0.84-0.96) & 0.93 (0.90-0.96) & 0.85 (0.81-0.89) & \textbf{0.99 (0.98-1.00)} \\
GONet (DINOv2 + MSD) & \textbf{0.85 (0.84-0.85)} & \textbf{0.87 (0.84-0.90)} & \textbf{0.90 (0.86-0.94)} & \textbf{0.93 (0.87-0.97)} & 0.93 (0.92-0.94) & \textbf{0.89 (0.86-0.92)} & \textbf{0.99 (0.99-1.00)} \\ \bottomrule
\end{tabular}%
} \captionsetup{labelformat=empty}
\captionof{table}{AUC values for eight different diagnostic models. The baselines are calculated using exclusively the rim-to-disc ration (RDR) and cup-to-disc ratio (CDR). The six pre-trained Vision transformer models are fine-tuned solely on the KULRD-Train dataset, whereas GONet uses a DINOv2 backbone and employs a multi-source domain (MSD) leave one domain out fine-tuning strategy. The parentheses report 95\% confidence intervals (CI).}
\label{tab:base-results}
\end{figure*}
\subsubsection{GONet}
The selected pre-trained model was fine-tuned to the downstream binary classification task of GON identification. Similarly to the work of Men et al. \cite{Men2023DRStageNet:Images}, a multisource domain training (MSD) approach was experimented with. This approach involves performance of the fine-tuning step on a joint set of multiple-source datasets while evaluating generalization performance on a single left-out target domain. The rationale behind this approach is that when training a model on a single dataset, it may overfit to this specific domain distribution. KULRD was divided into KULRD-Train, which was included in all experiments, and KULRD-Test. At each fine-tuning stage, a joint training dataset was used, consisting of 90\% of all source domains, except the left-out target domain for which model performance is reported. Similarly, a joint validation dataset, consisting of the remaining 10\% DFIs, was used. Data augmentation, including brightness changes, zoom changes, rotation and horizontal and vertical flips, was performed during the fine-tuning stage. The hyperparameters chosen for this task were identical to those of Men et al. \cite{Men2023DRStageNet:Images}, due to the similarity of the task. The resulting model is denoted GONet.

\begin{figure}[h]
\noindent\textbf{Table III: Comparison between GONet and current SOTA}
\vspace{0.5em}

\resizebox{\columnwidth}{!}{%
\begin{tabular}{@{}cccccc@{}}
\toprule
\textbf{Dataset} & \multicolumn{1}{l}{\textbf{\#DFIs}} & \multicolumn{1}{l}{\textbf{GON\%}} & \textbf{\begin{tabular}[c]{@{}c@{}}Current SOTA\\ (95\% CI if available)\end{tabular}} & \textbf{\begin{tabular}[c]{@{}c@{}}CDR\\ (95\% CI)\end{tabular}} & \textbf{\begin{tabular}[c]{@{}c@{}}GONet\\ (95\% CI)\end{tabular}} \\ \midrule
DRISHTI-GS & 101 & 69\% & 0.92 \cite{Sreng2020DeepImages} & 0.87 (0.78-0.95) & \textbf{0.94 (0.88-0.98)} \\
REFUGE & 1,200 & 10\% & 0.95 (0.92-0.98) \cite{Hemelings2023AImages}& \textbf{0.96 (0.94-0.98)} & 0.93 (0.90-0.96) \\
REFUGE2-Val & 400 & 25\% & 0.91 \cite{Hemelings2023AImages} & 0.90 (0.85-0.94) & \textbf{0.94 (0.90-0.96)} \\
REFUGE2-Test & 400 & 25\% & 0.87 \cite{Hemelings2023AImages} & 0.81 (0.75-0.86) & \textbf{0.88 (0.82-0.92)} \\
GAMMA-Train & 100 & 50\% & 0.99 (0.97-1.00) \cite{Hemelings2023AImages} & 0.92 (0.85-0.987) & \textbf{0.99 (0.98-1.00)} \\ \bottomrule
\end{tabular}%
} \captionsetup{labelformat=empty}
\captionof{table}{Comparison between GONet and current state-of-the-art (SOTA) reporting out-of-distribution (OOD) generalization on open-access datasets used in our experiments. To perform a strict comparison between GONet and other works, no exclusion criteria were applied to the datasets.}
\label{tab:sota-comparison}
\end{figure}

\subsubsection{Benchmarks}
We evaluated the advantage of using DL with raw DFI versus disc derived features only. More specifically, GONet was compared to the standard clinical CDR measures as well as the RDR \cite{Kumar2019Rim-to-DiscPrescreening} estimated using LUNet (see Section \ref{subsec:lunet}). In addition, we reviewed the recent literature relating to GON diagnosis from DFI and identified research reporting OOD generalization performance on datasets included in our research, and report those results as benchmarks to ours. The performance with REFUGE, REFUGE2 and GAMMA was reported by Hemelings et al. \cite{Hemelings2023AImages}, and with DRISHTI-GS was reported by Sreng et al. \cite{Sreng2020DeepImages}. Hemelings reported the performance with two subsets of REFUGE2 containing 400 DFIs each, as well as a subset of GAMMA, containing 100 DFIs. To strictly compare between their reported results and ours, we evaluated the performance of GONet on these datasets without applying any exclusion criteria. Finally, we compared GONet to the open-source model Brighteye \cite{Lin2024Brighteye:Transformer}, which is a ViT-based method trained using a dataset of 101,442 silver-standard annotated DFIs.

\subsubsection{Performance measures}
\label{subsec:performance_measures}
AUC was computed with a confidence interval obtained by bootstrapping 1000 times 95\% of the test set. In addition, Wilcoxon-signed-rank test \cite{Rey2011Wilcoxon-Signed-RankTest} was used to show a statistical difference in performance between GONet and other benchmark models.
Finally, to measure the generalizability and prediction quality of GONet, the Brier score was used \cite{Brier1950VerificationProbability}.

%%
%\subsubsection{Error analysis}
%We examined instances where GONet misclassified DFIs. Each DFI underwent an independent and blind review by two GON specialists (HP and EZB) who were unaware of the dataset origin and the assigned reference label. The specialists assessed whether the DFIs were ungradable, identified the presence of one or more comorbidities graded them as GON+ and GON-. Disagreements between the annotations of the two specialists were discussed, and a consensus was reached. For false negative the CDR was computed and compared to the one of the true positive.

%%%%%%%%%%%%%%%%%%%%%%%%%%%%%
\section{Results}

%%%%%%%%%%%%
\begin{figure*}[h]
    \centering
    %\hspace{-1.8cm}
    \includegraphics[width=1\textwidth]{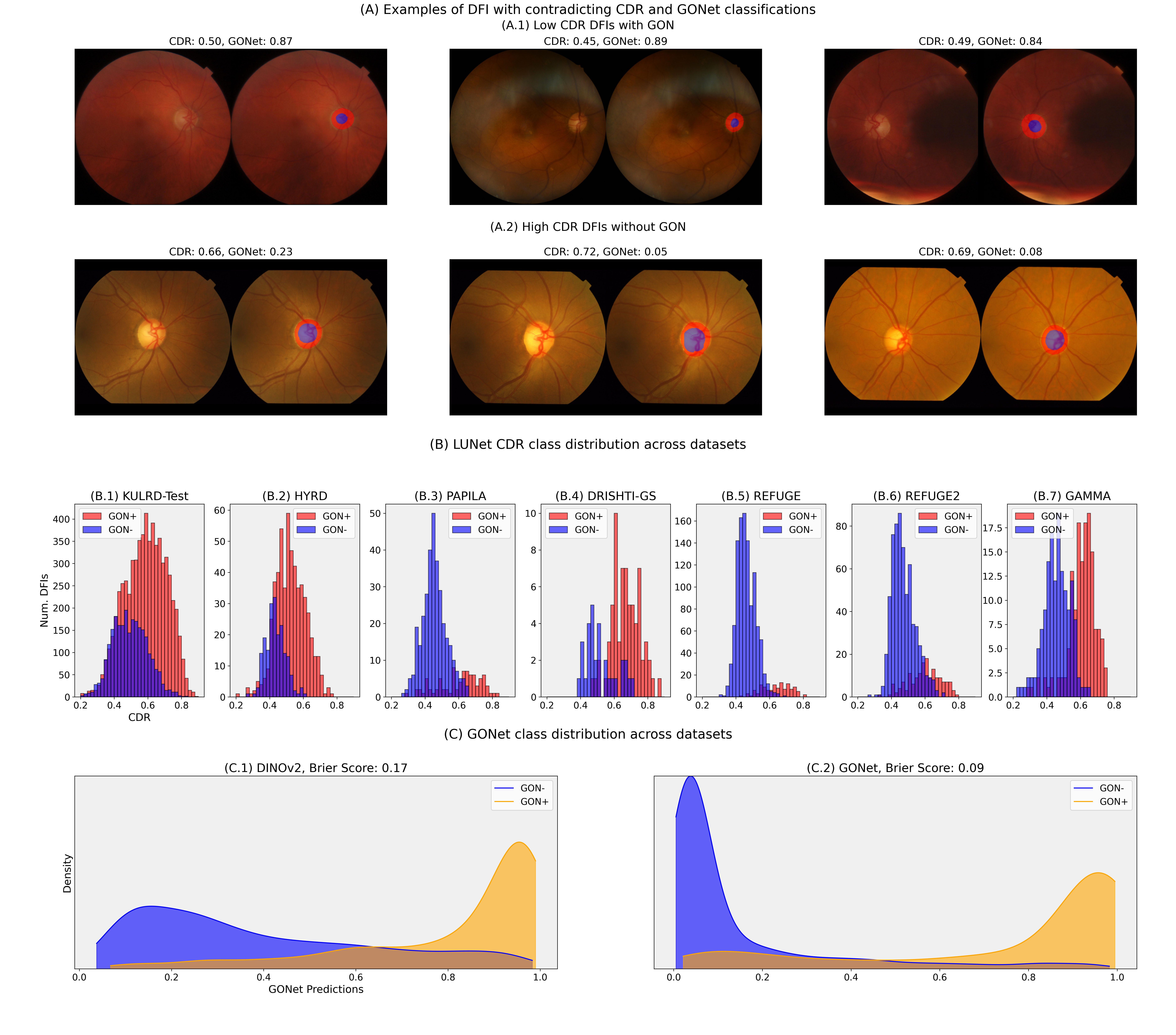}
    \caption{Panel A: Examples of DFIs with low CDR ($\leq$0.5) that are GON+ (A.1) and DFIs with high CDR ($\geq$0.65) that are GON-, and which were identified as such with certainty of $\geq$0.75. Panel B: CDR distribution of GON+ and GON- DFIs per dataset. Panel C: Distribution of DINOv2 (C.1) and GONet (C.2) predictions for GON+ and GON- DFIs over all datasets, using kernel density estimation (KDE). Brier score \cite{Brier1950VerificationProbability} is reported for each model.}
    \label{fig:generalization-results}
\end{figure*}

%%%%%
\subsection{GONet}
The DINOv2 pre-trained ViT outperformed other pre-trained models on the KULRD-Test set, achieving an AUC of 0.84 (0.83-0.85) on the KULRD-Test set (Figure \ref{fig:main_results}B). The RETFound approach \cite{Zhou2023AImages}, which was pre-trained in-domain with 900K DFIs, exhibited comparable performance, with an AUC of 0.83 (0.82-0.84). Other SSL methods, including base ViT-B and Swin Transformer, achieved lower performance. Based on this, DINOv2 was selected as the base model and further trained using the MSD approach. GONet significantly ($p<0.05$) outperformed the SSD in terms of AUC, across all datasets (Figure \ref{fig:main_results}C). Numerical results are also reported in Table \ref{tab:base-results}. The AUC of GONet over the target domains was within the range of 0.85 and 0.99.  Panel C of Figure \ref{fig:generalization-results} presents the density histogram of DINOv2 and GONet predictions. The Brier score for GONet is 0.09, which is nearly twice as small as the score of 0.17 for DINOv2. 

%%%%%
\subsection{GONet vs. disc features}
LUNet achieved a MAE of 0.043 in estimating CDR on the REFUGE dataset (Figure \ref{fig:main_results}A). For comparison, a recent study \cite{Gao2024AutomatedDetection} reported an MAE of 0.043 on REFUGE when considering the dataset as a target domain. Thus, the estimation of CDR using LUNet is in congruence with a recent report and constitutes a fair baseline. GONet outperformed CDR in six out of seven domains, with up to 21.6\% improvement in AUC (Figure \ref{fig:main_results}D). The performance of GONet was inferior to CDR on the REFUGE dataset only. RDR underperformed both GONet and CDR across all target domains. 

% 0.035 after fine-tuning their model using 800 DFIs and subsequently testing on a distinct subset of 400 DFIs in REFUGE. Without performing fine-tuning on a REFUGE subset the authors obtained an MAE of 

%%%%%
\subsection{GONet vs. other research works}
GONet outperformed Brighteye in six out of seven domains, achieving an improvement of up to 12.9\% in AUC (Figure \ref{fig:main_results}D). GONet and Brighteye performance was comparable for GAMMA, with an AUC of 0.99. When comparing GONet to published results reporting on OOD generalization on some of the open datasets used for our experiments, GONet improved over SOTA \cite{Sreng2020DeepImages,Hemelings2023AImages} in four out of five instances (Table \ref{tab:sota-comparison}). The improvement ranged from 0.3\% (0.987 vs. 0.990 for GAMMA-Train) to 2.6\% for Drishti-GS. For the REFUGE dataset, the best performance was obtained using CDR.

%%%%%%%%%%%%
%\subsection{GONet generalization}

%%%%%%%%%%%
%\subsection{Error analysis}
%TODO

%%%%%%%%%%%
\section{Discussion}
In this research, we evaluated six different pre-trained vision transformers and four SSL methods for the task of GON identification. The DINOv2 backbone exhibited superior performance in the source domain test set (Figure \ref{fig:main_results}B). Notably, the performance of RETFound, a DFI-based foundation model, was not superior to DINOv2, which is a foundation model trained using natural images. This suggests that use of a large number of in-domain images, as in RETFound, does not provide an advantage over use of a very large number of natural images for pretraining with SSL. When using multi-source domain fine-tuning, GONet exhibited significantly better generalization performance than when training on a single source (Figure \ref{fig:main_results}C, Figure \ref{fig:generalization-results}C). These results, both visually and quantitatively highlight the superior generalization performance of GONet over the SSD approach. Specifically, they highlight the value of using MSD in learning a more generalizable representation for a given task, as it avoids overfitting a specific domain or learning shortcut features. 

An important element of this work was the comparison between the usage of CDR versus raw DFI as input for a DL model. The added value of using GONet over CDR confirms and quantifies the benefit of utilizing the entire DFI, indicating that GON can influence DFI beyond the appearances of the OD and OC \cite{Hemelings2021DeepDisc}. GON may cause structural changes in the retinal nerve fiber layer and other retinal layers, which may manifest as alterations in the retinal vasculature and localized thinning of the retina, which could influence the appearance of areas beyond the optic nerve head \cite{Fhima2024ComputerizedGlaucoma, Hemelings2021DeepDisc}. Figure \ref{fig:generalization-results}A displays cases where the CDR failed to capture GON, while GONet was successful.

Although CDR provided reasonable results across external target domains, there was a discrepancy in its performance on KULRD-Test, where it led to a low AUC of 0.72. Figure \ref{fig:generalization-results}B illustrates the CDR distribution, as estimated using LUNet, across all datasets used in this study. Notably, for each patient group, the CDR appeared to follow a normal distribution centered around the median CDR, ranging from 0.43 to 0.52 (95\% CI) for GON- patients and from 0.52 to 0.67 (95\% CI) for GON+ patients. The absence of such a clear division in KULRD-Test, unlike in the other datasets, may be due to KULRD including a broader range of conditions that affect the OD, such as optic disc drusen and papilledema. In contrast, public datasets are selective and typically provide biased datasets, including selected patient profiles which may not represents the intended population sample. Another hint to dataset bias can be seen in the number of DFIs excluded because of low quality or missing OD. For HYRD, it was 15.7\%. For datasets used in competitions, such as REFUGE, REFUGE2 and GAMMA, less than 0.5\% of the DFIs were filtered out, suggesting a pre-selection of the DFIs included in the competition.
%Another explanation may be the relatively large proportion of patients with normal tension glaucoma, which often have very focal rim defects, resulting in lower CDR despite clearly being glaucomatous. 

A portion of DFIs from KULRD were automatically excluded (13.9\%, Figure \ref{fig:eligibility-flowchart}), primarily due to poor image quality. In the context of clinical deployment, this suggests that technicians capturing DFIs would be prompted in real time by our system to retake images deemed of insufficient quality. This process is consistent with current clinical practices, where technicians typically capture multiple images per eye to ensure that at least some meet the quality standards required for interpretation by retinal specialists.

This research has several limitations. Despite the relatively large number of datasets included in the study, the model should be evaluated on additional datasets from medical centers around the world, to cover a wider range of ethnicities, comorbidities, medical center practices, and camera types and FOVs. These datasets should accurately reflect real life, as currently four of the five SOTA results we compared GONet to belong REFUGE, REFUGE2 and GAMMA, which are very biased. Additionally, a set of common pre-trained Vision transformers was evaluated in this research. A more exhaustive benchmark of all SOTA vision transformers may be beneficial. 

Overall, this research introduced a novel DL model, GONet, designed for GON identification from single DFI images. GONet demonstrates robustness, achieving high performance and strong OOD generalization, as evidenced across seven datasets. Additionally, this work contributed to the field by providing HYRD, a new dataset we developed, as an open-access resource. GONet is accessible at [URL upon publication.]

%%%%
\subsection*{Data availability}
All datasets used in our experiments are open-access, except for KULRD. They can be found at the following URLs: \href{https://www.kaggle.com/datasets/lokeshsaipureddi/drishtigs-retina-dataset-for-onh-segmentation}{DRISHTI-GS}, \href{https://figshare.com/articles/dataset/PAPILA/14798004/1?file=28454352}{PAPILA}, \href{https://www.kaggle.com/datasets/arnavjain1/glaucoma-datasets?resource=download}{REFUGE}, \href{https://www.kaggle.com/datasets/arnavjain1/glaucoma-datasets?resource=download}{ORIGA}, \href{https://www.kaggle.com/datasets/arnavjain1/glaucoma-datasets?resource=download}{G1020}, \href{http://hdmilab.cn/ichallenge}{REFUGE2}, \href{http://hdmilab.cn/ichallenge}{GAMMA}. HYRD is a new open-access dataset contributed by the present research [URL provided on publication.]).

\printbibliography

\end{document}